# Study of dopant concentrations on thermal induced mode instability in high power fiber amplifiers

Rumao Tao, Pengfei Ma, Xiaolin Wang, Pu Zhou, Zejin Liu

*Abstract*—Dependence of mode instabilities (MI) on ytterbium dopant concentrations in high power fiber amplifiers has been investigated. It is theoretically shown that, by only varying the fiber length to maintain the same total small-signal pump absorption, the MI threshold is independent of dopant concentration. MI thresholds of gain fibers with ytterbium dopant concentration of $5.93 \times 10^{25}/m^3$ and $1.02 \times 10^{26}/m^3$ have been measured, which exhibit similar thresholds and agree with theoretical results. The result indicates that heavy doping of active fiber can be adopted to suppress nonlinear effects without decreasing MI threshold, which provides a method of maximizing the power output of fiber laser, taking into account the stimulated Brillouin scattering, stimulated Raman Scattering, and MI thresholds simultaneously.

## I. INTRODUCTION

TREMENDOUS progress in the area of high power fiber laser systems with near diffraction limited beam quality has been achieved in the past few years, which is due to their numerous applications, especially in coherent lidar system, nonlinear frequency conversion, coherent beam combining [1-4]. However, power scaling of ytterbium doped fiber laser systems with near diffraction limited beam quality is now limited by the sudden onset of mode instabilities (MI), which deteriorates the beam quality and pointing stability [5, 6], and limits the application of fiber lasers in the aforementioned areas. Concerning the rapid power scaling of fiber lasers in the mid-infrared spectral range [7, 8], it may be only a matter of time until MI also becomes a problem for other rare-earth-doped fiber lasers [9], which means that MI is one more fundamentally nonlinear challenging aspect for the power scaling capabilities of fiber laser. Due to the far-reaching impact of MI, lots of work has been carried out to gain a deep insight into this phenomenon, in which influence of various fiber parameters, i.e. core diameter [10-12], core *NA* [13], shape of dopant area [14-17], pump cladding diameter [11, 12, 15], have been studied. Dopant concentration of ytterbium ions is an important parameter of fiber. By increasing the dopant concentration, the length of gain fiber can be shortened, which has advantage in suppressing length-related nonlinear effects, such as Stimulated Brillouin Scattering (SBS) and Stimulated Raman Scattering (SRS)[18]. Nevertheless, the effects of ytterbium dopant concentrations on MI have not been fully studied except for some numerical investigation [19].

In this paper, dependence of MI on ytterbium dopant concentrations in high power fiber amplifiers has been investigated. Based on a semi-analytical model, the influence of dopant concentrations on MI has been studied and discussed. Then we built up a high power master oscillator power amplifier, and two fibers with different dopant concentrations were employed to verify the theoretical results. Agreement between theoretical predication and experimental results has been achieved.

## II. NUMERICAL STUDY

The fundamental mechanism leading to MI is thought to be stimulated thermal Rayleigh scattering [11, 20], which stems from the creation of a thermally-induced index grating written in the fiber by modal interference [5, 21], and has been employed widely in numerical investigation. Based on the aforementioned physical principle, the fraction of HOM for the case that MI is seeded by intensity noise of the signal laser can be expressed as [12, 22]

$$\xi(L) \approx \xi_0 \exp\left[\int_0^L dz \iint g(r,\phi,z)(\psi_2\psi_2 - \psi_1\psi_1) r dr d\phi\right] \\ \times \left\{1 + \frac{1}{4}\sqrt{\frac{2\pi}{\int_0^L P_1(z)|\chi''(\Omega_0)|dz}} R_N(\Omega_0) \exp\left[\int_0^L P_1(z)\chi(\Omega_0) dz\right]\right\} \quad (1)$$

where $R_N(\Omega)$ is the relative intensity noise of the input signal, $\xi_0$ is the initial HOM content, $\chi(\Omega)$ is the nonlinear mode coupling coefficient, which can be expressed as

$$\chi(\Omega) = 2\frac{n_0 \omega_2^2}{c^2 \beta_2} \mathrm{Im}\left(4 n_0 \varepsilon_0 c \iint \bar{h}_{12} \psi_1 \psi_2 r dr d\phi\right) \quad (2)$$

Here, only heat resulting from quantum defect has been considered. It is shown that photodarkening has significant impact on MI [23-25]. However, the origin and physical causes of the phenomenon are still under debate, and benchmarking techniques have been developed to mitigate it [26], which can also be employed to suppress its effect on MI [23, 27]. So the model has not taken it into consideration.

Other parameters are given as

$$\bar{h}_{kl}(r,\phi,z) = \frac{\eta}{\pi \rho C}\left(\frac{v_p - v_s}{v_s}\right) \sum_{\nu}\sum_{m=1}^{\infty} \frac{R_\nu(\delta_m, r) B_{kl}(\phi,z)}{N(\delta_m)\alpha\delta_m^2 - j\Omega} \quad (3a)$$

---

Rumao Tao, Pengfei Ma, Xiaolin Wang, Pu Zhou, Zejin Liu are all with College of Optoelectric Science and Engineering, National University of Defense Technology, Changsha, Hunan 410073, China (e-mail: taorumao@gmail.com).

$$B_{kl}(\phi, z) = \tag{3b}$$
$$\int_0^{2\pi} d\phi' \int_0^R g_0 R_\nu(\delta_m, r') \cos\nu(\phi-\phi') \frac{\psi_k(r',\phi')\psi_l(r',\phi')}{(1+I_0/I_{saturation})^2} dr'$$

$$g_0 = \frac{P_p(z)(\sigma_p^a \sigma_s^e - \sigma_p^e \sigma_s^a)/h\nu_p A_p - \sigma_s^a/\tau}{P_p(z)(\sigma_p^a + \sigma_p^e)/h\nu_p A_p + 1/\tau} N_{Yb}(r,\phi) \tag{3c}$$

$$I_{saturation} = \left[P_p(z)(\sigma_p^a + \sigma_p^e)/h\nu_p A_p + 1/\tau\right] \frac{h\nu_s}{\sigma_s^a + \sigma_s^e} \tag{3d}$$

$$N(\delta_m) = \int_0^R r R_\nu^2(\delta_m, r) dr \tag{3e}$$

Here $\Omega = \omega_1 - \omega_2$, $\nu_{p(s)}$ is the optical frequencies, $\eta$ is the thermal-optic coefficient, $\rho$ is the density, $C$ is the specific heat capacity, and $R$ is the radius of the pump cladding, $g(r, \phi, z)$ is the gain distribution in fiber and $\psi_2(r, \phi)$ and $\beta_2$ is the normalized mode profiles and propagation constant of high order mode (HOM) (LP$_{11}$ in the paper). $R_\nu(\delta_m, r) = J_\nu(\delta_m, r)$ ($J_m$ represents Bessel functions of the first kind) and $\delta_m$ is the positive roots of $\delta_m J_\nu'(\delta_m R) + h_q/\kappa J_\nu(\delta_m R) = 0$ ($h_q$ is the convection coefficient for the cooling fluid and $\kappa$ is the thermal conductivity). In Eq. (1), the first term on the right hand corresponds to the amplification of HOM by the laser gain while the second term corresponds to the amplification by the nonlinear mode coupling due to stimulated thermal Rayleigh scattering.

Based on Eq. (1), we calculated the threshold under different dopant concentration for different pump configurations, which is shown in Fig. 1. The fiber parameters are listed in Table I. All the fiber is fully doped. As dopant concentration of the fiber varies, other parameters remains unchanged except the fiber length, which is adjusted accordingly to the minimum value necessary to achieve high efficiency, defined as total small-signal pump absorption of 13dB. It shows from Fig. 1 that the MI threshold is independent of dopant concentration regardless of the pump schemes. Here, the counter-pumped case has lower threshold than that for co-pumped case, which seems different from the results in [28]. This is caused by the difference of fiber parameters, which results in different gain saturation and leads to the difference in these cases [11].

TABLE I PARAMETERS OF TEST AMPLIFIER

| | | | |
|---|---|---|---|
| $R_{core}$ | 15μm | $N_{Yb}$ | varies |
| $R$ | 125μm | $\sigma_p^a$ | $2.47\times10^{-24}$ m$^2$ |
| $n_{clad}$ | 1.45 | $\sigma_p^e$ | $2.44\times10^{-24}$ m$^2$ |
| $NA$ | 0.065 | $\sigma_s^a$ | $6.0\times10^{-27}$ m$^2$ |
| $\lambda_p$ | 976nm | $\sigma_s^e$ | $3.58\times10^{-25}$ m$^2$ |
| $\lambda_s$ | 1064nm | $\tau$ | 901 μs |
| $h_q$ | 5000 W/(m$^2$K) | $R_N$ | $10^{-10}$ |
| $\eta$ | $1.2\times10^{-5}$ K$^{-1}$ | $\xi_0$ | 0.01 |
| $\kappa$ | 1.38 W/(Km) | $P_l(0)$ | 10W |
| $\rho C$ | $1.54\times10^6$ J/(Km$^3$) | | |

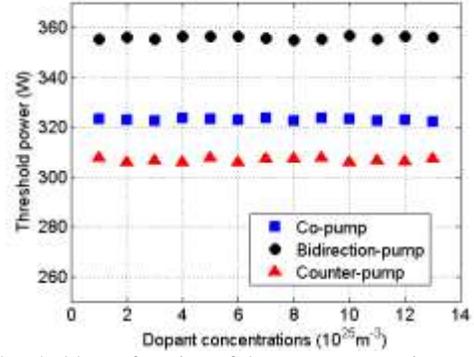

Fig. 1 Threshold as a function of dopant concentrations.

Due to that the second term on the right hand in Eq. (1) corresponds to the amplification by the nonlinear mode coupling, this term was analyzed in specific to explain the above theoretical results. By normalizing the fiber length to 1, the second term can be rewritten as

$$1 + \frac{1}{4}\sqrt{\frac{2\pi}{(N_{Yb}L)\int_0^1 P_l(L\mathbf{Z})|\chi''(\Omega_0)|d\mathbf{Z}}} \tag{4}$$
$$\times R_N(\Omega_0) \exp\left[(N_{Yb}L)\int_0^1 P_l(L\mathbf{Z})\chi(\Omega_0)d\mathbf{Z}\right]$$

with

$$B_{kl}(\phi, L\mathbf{Z}) = \tag{5a}$$
$$\int_0^{2\pi} d\phi' \int_{R_{dop}} \bar{g}_0 R_\nu(\delta_m, r') \cos\nu(\phi-\phi') \frac{\psi_k(r',\phi')\psi_l(r',\phi')}{(1+I_0/I_{saturation})^2} dr'$$

$$\bar{g}_0 = \frac{P_p(L\mathbf{Z})(\sigma_p^a \sigma_s^e - \sigma_p^e \sigma_s^a)/h\nu_p A_p - \sigma_s^a/\tau}{P_p(L\mathbf{Z})(\sigma_p^a + \sigma_p^e)/h\nu_p A_p + 1/\tau} \tag{5b}$$

$$I_{saturation} = \left[P_p(L\mathbf{Z})(\sigma_p^a + \sigma_p^e)/h\nu_p A_p + 1/\tau\right] \frac{h\nu_s}{\sigma_s^a + \sigma_s^e} \tag{5c}$$

where $R_{dop}$ is the radius of the doping area, and the other parameters are the same as in Eq. (3). By adapting the fiber length accordingly, the total small-signal pump absorption keeps the same as dopant concentration changes, which results in that the term ($N_{Yb}L$) in Eq. (4) is unchanged. On the other hand, although the dopant concentration has changed, the pump/signal power distribution along the normalized fiber length remains the same as shown in Fig. 2(a), which results that $\bar{g}_0$ and $I_{saturation}$ are the same, and ultimately causes that the nonlinear mode coupling coefficients at different normalized fiber lengths keep the same. The nonlinear mode coupling coefficients were plotted in Fig. 2(b), which were obtained by the combination of Eq. (2) and Eq. (3) with Eq. (3b) and (3c) replaced by Eq. (5). The nonlinear mode coupling coefficients is far smaller than those in [16], which is due to that the term $N_{Yb}$ has been extracted in Eq. 4. It revealed in Fig. 2 that, although the dopant concentrations in the fiber are different, the nonlinear mode coupling coefficients are the same at the same relative length. Finally, we can achieve that the value of Eq. (4) is independent of the dopant concentrations, which results in that, with the other parameters keeping constant and the fiber length being adapted to maintain the same total small-signal pump absorption, the variation of dopant concentrations has no impact on MI.

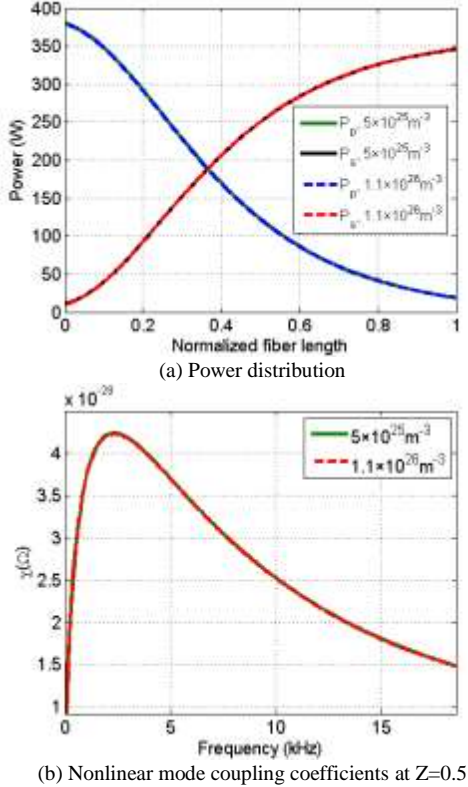

(a) Power distribution

(b) Nonlinear mode coupling coefficients at Z=0.5

Fig. 2 Power distribution and nonlinear mode coupling coefficients at different dopant concentrations for co-pump schemes.

Based on the aforementioned study, we can also obtain that the length of fiber has no impact on the threshold of MI as long as that only the dopant concentration is varied to maintain the same total small-signal pump absorption, which is show in Fig. 3(a). This is not in contrast with the results in [29], in which the pump cladding was adapted to maintain the same total small-signal pump absorption and the dopant concentrations was kept the same. Similar to the case in [29], we can obtain similar results as shown in Fig. 3(b). As the diameter of pump cladding increase (shown in Fig. 3(b)), the gain saturation becomes stronger, which results in MI threshold increasing. Meanwhile, longer fiber is needed for larger pump cladding diameter to guarantee efficient pump absorption. Then the MI threshold seems related to the fiber length.

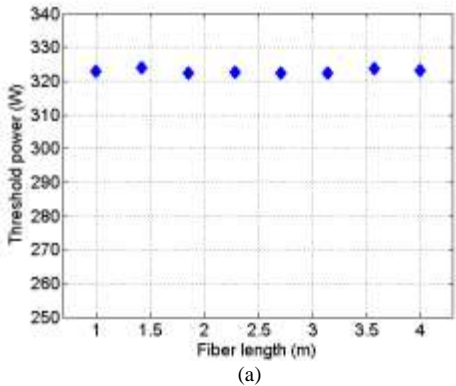

(a)

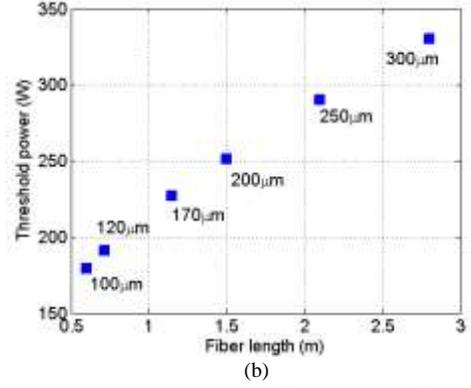

(b)

Fig. 3 Nonlinear mode coupling coefficients at different normalized fiber lengths.

## III. EXPERIMENTAL RESULTS AND DISCUSSIONS

To validate the aforementioned prediction, experimental investigation has been performed. The experiment setup is shown in Fig. 4. The main amplifier employed large mode area (LMA) ytterbium-doped fiber (YDF) with core diameter being 30μm and clad diameter being 250μm, which is seeded by a 1080nm seed with ~10W power, and pumped in the co-propagating direction by fiber pigtailed laser diodes (LD) at 976nm. A home-made cladding mode striper (CMS) was employed to strip the residual pump laser and cladding mode. The output end of the delivery fiber is angle cleaved at 8°. The onset of MI was monitored by detecting the time fluctuation of scattering power with photo-detector [30]. Gain fibers with different ytterbium dopant concentrations have been employed in the experiment, and the parameter are listed in table II. Higher dopant concentration leads to the core NA of fiber B is larger than fiber A, but the influence of core NA on MI is negligible for the cases here [13].

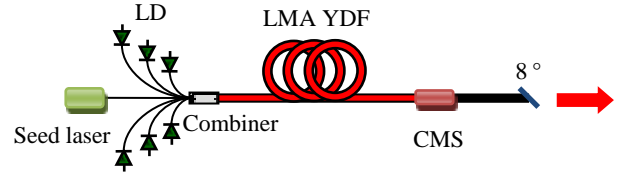

Fig. 4 Experimental setup of the high power fiber amplifier.

TABLE II PARAMETERS OF THE ACTIVE FIBER

|  | Core $NA$ | Dopant concentration |
|---|---|---|
| Fiber A | 0.065 | $5.93 \times 10^{25}/m^3$ |
| Fiber B | 0.07 | $1.02 \times 10^{26}/m^3$ |

The measured thresholds are 440W for fiber A and 446W for fiber B, respectively. To avoid the impact of photodarkening on the results, the MI thresholds are the power when MI was first encountered. It can be seen that, although the dopant concentration of ytterbium ions for fiber B is about two times of that for fiber A, they have similar MI threshold, which agrees with our previous theoretical results.

Based on the theoretical study in previous section and the experimental study in this sections, the dependence of MI on dopant concentrations has been investigated, which leads to an interesting finding: dopant concentration has little impact on the threshold of MI as long as that only the fiber length is varied to maintain the same total small-signal pump absorption. Thus, higher dopant concentrations can be adopted without decreasing the threshold of MI, which may find useful experimental and theoretical applications.

Various effective methods have been proposed to mitigate or suppress MI, such as tailing the Yb-ion concentration [14, 15, 31], shifting the pump wavelength [22, 32, 33], increasing the pump cladding diameter [11, 15]. In these methods, they share one aspect in common, which is that they all leads to reducing of the small-signal pump absorption. If the dopant concentrations keep the same, longer fiber is required to achieve high lasing efficiency, which inevitably leads to the detrimental effect on the SBS and SRS threshold. It seems that the suppression of MI is in conflict with suppression of SBS and SRS. However, with the conclusion achieved in the previous sections, such contradictions can be solved by increasing the dopant concentrations. In conjunction with the proposed MI suppression methods, higher doped fiber can realize the suppression of MI and SBS and SRS simultaneously [34]. Take the cases in [35] for example, in which the power scaling capacity of the amplifier was limited to 2.6kW by SRS, SRS threshold can be further increased by employing fiber with higher doping concentration while maintaining the merit of no MI. Nevertheless, in some method, suppression of MI is implemented through mode specific loss by tight coiling the fiber [12, 24, 36], in which longer length of fiber can provide higher high order mode loss and stronger MI suppression. When adopting higher doped fiber to suppress nonlinear effects in these cases, it should take cautions to not reduce the fiber length too much to make MI suppression invalidate.

Furthermore, the computation resources can be saved by increasing the dopant concentration to shorten the length of the fiber in the calculation of MI threshold, which is useful for some time consuming numerical models [37, 38].

## IV. CONCLUSIONS

In summary, we studied the effects of dopant concentrations on mode instabilities. Dependence of MI on ytterbium dopant concentrations has been investigated theoretically, which reveals that the MI threshold is independent of dopant concentration. Then the threshold of amplifiers using 30/250 fibers with dopant concentration of $5.93 \times 10^{25}/m^3$ and $1.02 \times 10^{26}/m^3$ has been experimentally examined. It shows that they have similar MI threshold. The experimental results agree with the theoretical predictions, which mean that dopant concentration has little impact on MI. The results may find useful applications in areas that need shorten the length of the fiber to suppress nonlinear effects, such as SBS and SRS, as well as to save the computation resources of numerical simulation.